\pdfoutput=1
\documentclass[english]{cccconf}
\usepackage{booktabs}
\usepackage{multirow}
\usepackage{threeparttable}
\usepackage{makecell}
\usepackage[comma,numbers,square,sort&compress]{natbib}
\usepackage{epstopdf}
\usepackage{cite}
\usepackage{amsmath}
\usepackage{amssymb}

\usepackage{subfigure}

\newtheorem{remark}{Remark}

\begin{document}

\title{Low-cost adaptive obstacle avoidance trajectory control for express delivery drone }
\author{Yanhui Zhang\aref{zju,nus},
       Caisheng Wei\aref{csu},
    Yifan Zhang\aref{zju},
    Congcong Tian\aref{nus,hit},
    Weifang Chen\aref{zju},
        }

\affiliation[zju]{School of Aeronautics and Astronautics, Zhejiang University, Hangzhou 310058, P.~R.~China
        \email{yanhuizhang@zju.edu.cn}}
\affiliation[nus]{Department of Electrical and Computer Engineering, National University of Singapore, Singapore 117583, Singapore}
\affiliation[csu]{School of Automation, Central South University, Changsha 410083, P.~R.~China}
\affiliation[hit]{School of Mechanical Engineering and Automation, Harbin Institute of Technology, Shenzhen, Guangdong 518055, P.~R.~China}
\maketitle

\begin{abstract}
This paper studies quadcopters' obstacle avoidance trajectory control (OATC) problem for express delivery. A new nonlinear adaptive learning controller that is low-cost and portable to different wheelbase sizes is proposed to adapt to large-angle maneuvers and load changes in UAV delivery missions. The controller consists of a nonlinear variable gain (NLVG) function and an extreme value search (ES) algorithm to reduce overshoot and settling time. Finally, simulations were conducted on a quadcopter to verify the effectiveness of the proposed control scheme under two typical collision-free trajectories.

\end{abstract}


\footnotetext{This work was supported in part by the National Natural Science Foundation of China for the Near Space Research Program under Grant U20B2007 and Grant 12002306. The first author of this paper (Yanhui Zhang) gratefully acknowledges the support by the China Scholarship Council (grant number 202306320387).}

\section{Introduction}
The obstacle avoidance trajectory control (OATC) problem of the uncrewed aerial vehicle (UAV) has received significant attention recently in the real world (indoor and outdoor) as illustrated in \citep{indoorQuadrotor1}. In these scenarios, the drones that can avoid obstacles have more application value for many missions, such as exploration, swarms flight, and prevention-control of new coronavirus (see \citep{CY_ZYH_TNNLS2023},  \citep{801SwarmsCollisionAvoidance}, and references therein). Since drones are not limited by infrastructure such as light poles, streams, and walls, they can lower labor costs and reduce package delivery time \citep{yanhuiXQZ_AST2023}. Significantly, the demand for drone delivery tasks is increasing sharply, and it is vital to realize obstacle avoidance flight control in 3D airlines \citep{13mansouri2020unified}. Hence, in the future scenario of dense air flight traffic, OATC between UAVs-UAVs and UAVs-MAVs (Manned Vehicle) will be sensitive.

In the last decades, several classical methodologies have been presented for OATC to achieve the flight control of quadcopters, such as detection and image processing in \citep{52yanhui2020CAC}. Also, for such UAV obstacle avoidance problems, some researchers have done some exciting work in the fields of motion planning for drones. Besides, some others focus on the controller problems, as our former work in \citep{ZYH2023TASE} and W. Bao described in \citep{BWM2023}. 
However, due to the varied layout and motors, quadcopters' parameters are different in practical scenarios. Hence, the controller transplant cost must be lowered to adapt to another flight platform. Thus, the NN controllers have yet to be used on a large scale in recent years. In recent years, some methods have been proposed to solve the problem (such as PID, ESC \citep{30yin2014SlideMode}). Some others designed the control system combined PID/MPC with robust control and learning-based adaptive control in \citep{ZYM2023TAES} and \citep{TrC2019CaishengWei_learnAdapCtrl}.
Motivated by the observation above, this paper proposes NLVG-PID, which could reduce the overshoot of fixed-gains PID controllers. The main contributions of this work lie in:
$\mathrm{1)}$ The structure of the fixed-gain PID controller and the nonlinear dynamics and kinematics models are analyzed in this work. 
$\mathrm{2)}$ The NLVG-PID controller of a delivery quadcopter is designed, and the extremum seeking is introduced to learn the optimal nonlinear PID parameter under boundary conditions. To our knowledge, this is the first NLVG-PID obstacle avoidance controller to be used in delivery drones.
$\mathrm{3)}$ A simulation was performed on a quadcopter to verify the effectiveness of the proposed control scheme under two typical collision-free trajectories. 

\begin{figure}[htbp]
	\centering
	\includegraphics[width=0.47\textwidth]{ 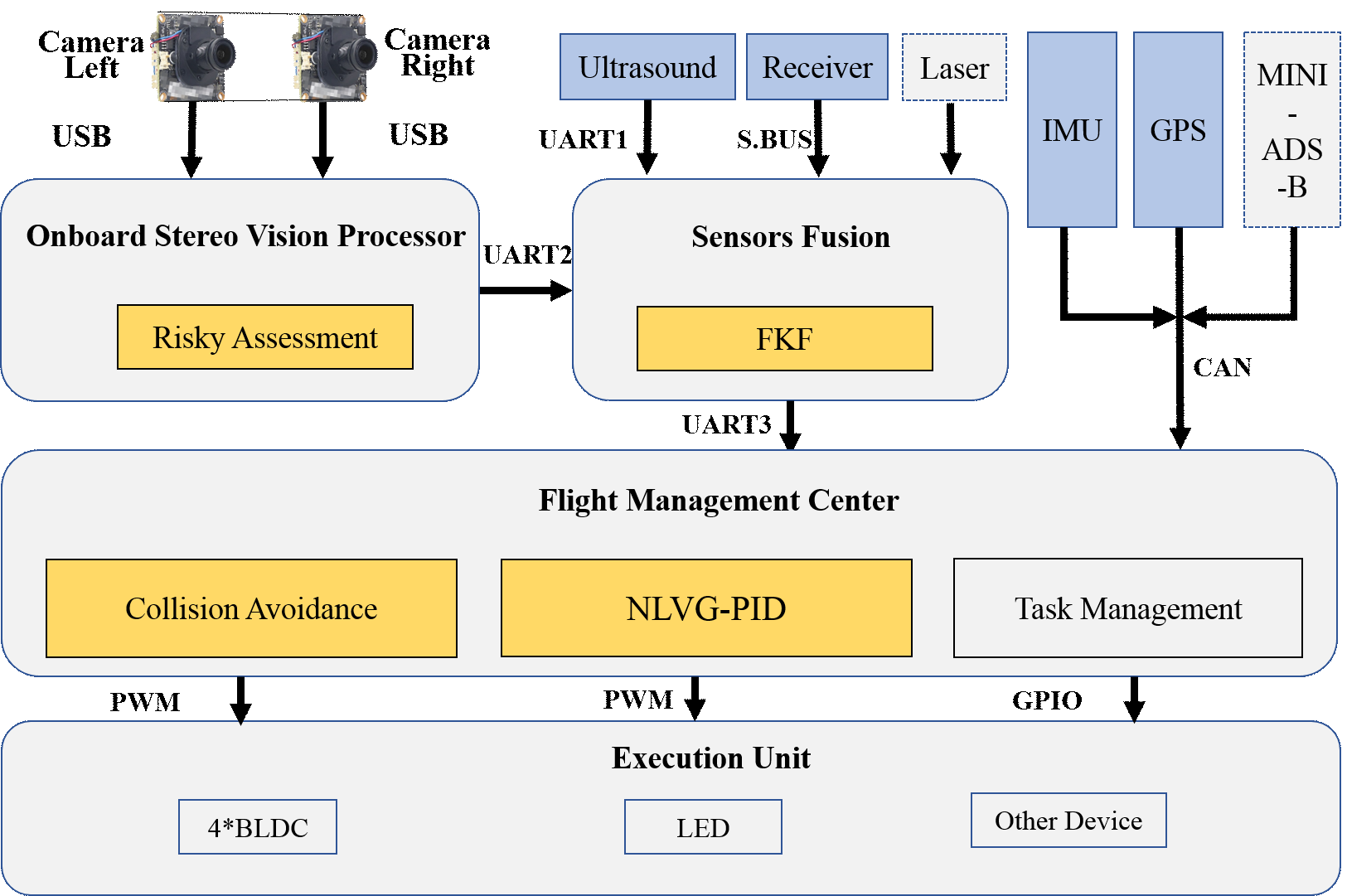}
	\caption{Scheme of NLVG-PID 3D delivery drone system.}
	\label{fig_Overview on our quadrotor system}
\end{figure}

The Scheme of quadcopter control system is described in Fig. \ref{fig_Overview on our quadrotor system}.

The remainder of this paper is organized as follows: Section \ref{Section_problem}, the mathematical model of dynamics, kinematics, and obstacle detection pipeline of the quadcopter are described in detail. In Section \ref{Sec_controller}, the NLVG-PID framework and an ES method for boundary optimal parameter determination are developed. The stability analysis of the closed-loop system is also provided in this section. In Section \ref{section_simulation}, the quadcopter's NLVG-PID controller system has been verified in four-side route, 8-character route, and scenic spiral position tracking, respectively. Section \ref{section_conclusion} concludes this work.

\section{Problem formulation}
\label{Section_problem}
\begin{figure}[htbp]
	\centerline
	{\includegraphics[width=0.47\textwidth]{ 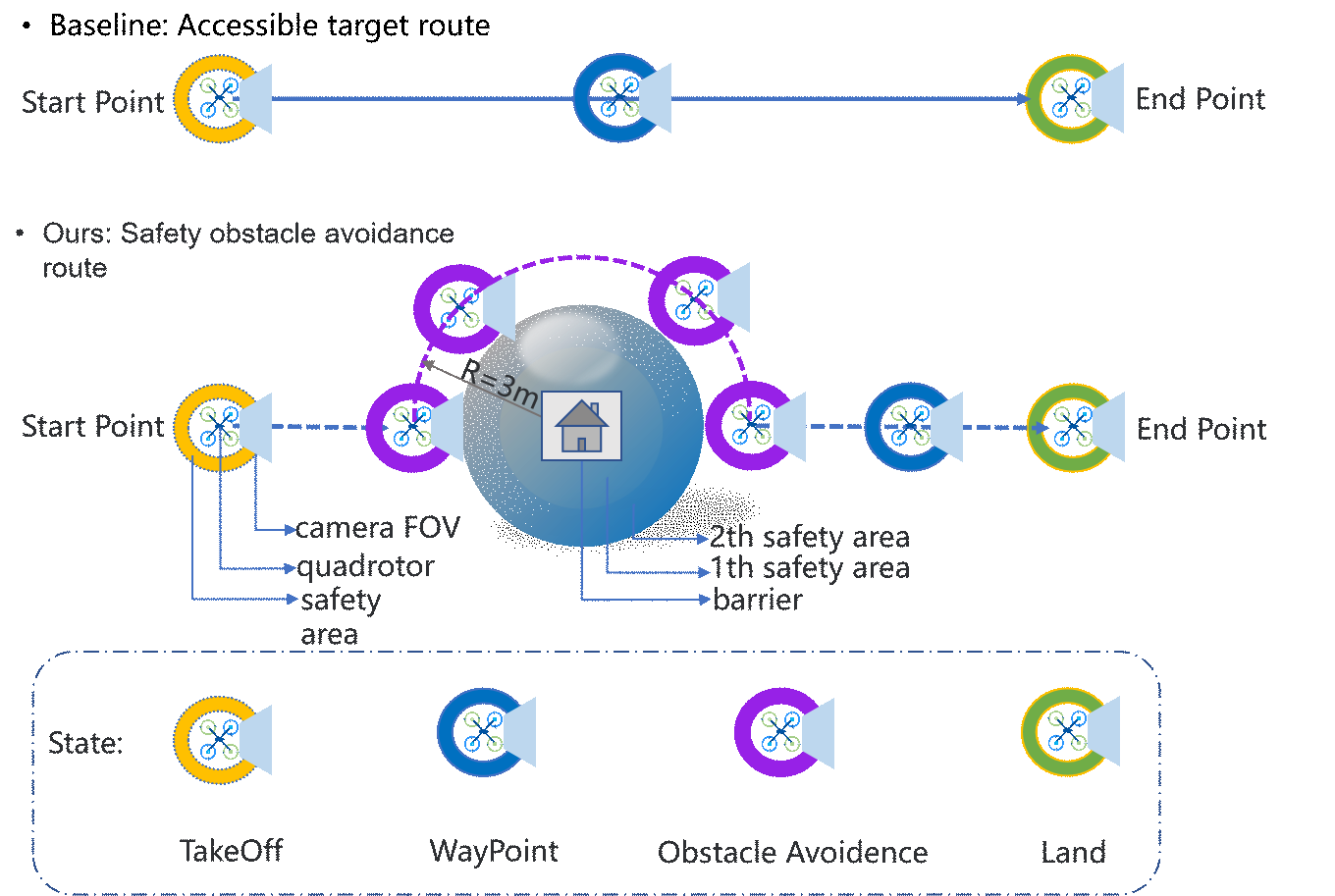}}
	\caption{Delivery drone 3D obstacle avoidance process.}
	\label{fig_3DobstacleAvoidanceProcess}
\end{figure}

This paper assumes that the structure of the express delivery drone, an X-type quadcopter from \citep{52yanhui2020CAC}, is known, but the specific load weight and the precise model of the drone are unknown. Obstacle avoidance is performed in three-dimensional space. First, the threat area is circled in the top view, as shown in Fig.\ref{fig_3DobstacleAvoidanceProcess}. Avoid obstacles, and then choose whether to fly upward or downward according to the distance cost in Euclidean space.

To facilitate the analysis of the rotation and translation of the UAV, a coordinate system is depicted as shown in our former work \citep{yanhui2022CCC}, the $B = (X_B, Y_B, Z_B)$ denotes the body-fixed coordinate system (B-frame), $E = (X_E, Y_E, Z_E)$ is the earth inertial frame (E-frame), $V_E = (v_{ex},v_{ey},v_{ez})$ indicates the velocity of quadcopter in E-frame.
The schematic diagram of the quadcopter control system includes the following components, as displayed in Fig. \ref{fig_Overview on our quadrotor system}.

\section{Controller design}\par
\label{Sec_controller}
Usually, PID controller is defined as follows:
\begin{equation} 
	U(t)_{j}^{\text{PID}}=K_{p}^{j} \cdot e(t)+K_{d}^{j}\cdot\dot{e}(t)+K_{i}^{j}\cdot \int_{0}^{t} e(\tau) d \tau ,
	\label{formula_PID_basic_law}
\end{equation}
with $j \in [x,y,z,\psi]$ and
\begin{equation}
\left\{
\begin{aligned}
	e(t) &= \zeta _{desire} - \zeta _{real},\\
    \dot{e}(t) &=\dot{ \zeta }_{desire} - \dot{\zeta }_{real}, 
\end{aligned}\right.
\end{equation}
$\zeta _{desire}$, $\zeta _{real}$ are the desired value and the real value of $\zeta $.

\begin{remark}
	It can be observed that the quadcopter system is highly nonlinear and strongly coupled.
\end{remark}

\begin{remark}
It can be seen that the quadcopter system is highly nonlinear and strongly coupled. Usually, in order to reduce the complexity of control system design, channel decoupling is used to decouple the three-axis position and attitude control.
\end{remark}

\subsection{NLVG-PID controller}
Defining nonlinear variable gains $f_{K_p(e)}$ of $P$ controller in channle $j$. Besides, $\delta_2$ impact on $f_{K_p(e)}$.
\begin{figure}[htbp]
	\centering
		\includegraphics[width=0.47\textwidth]{ 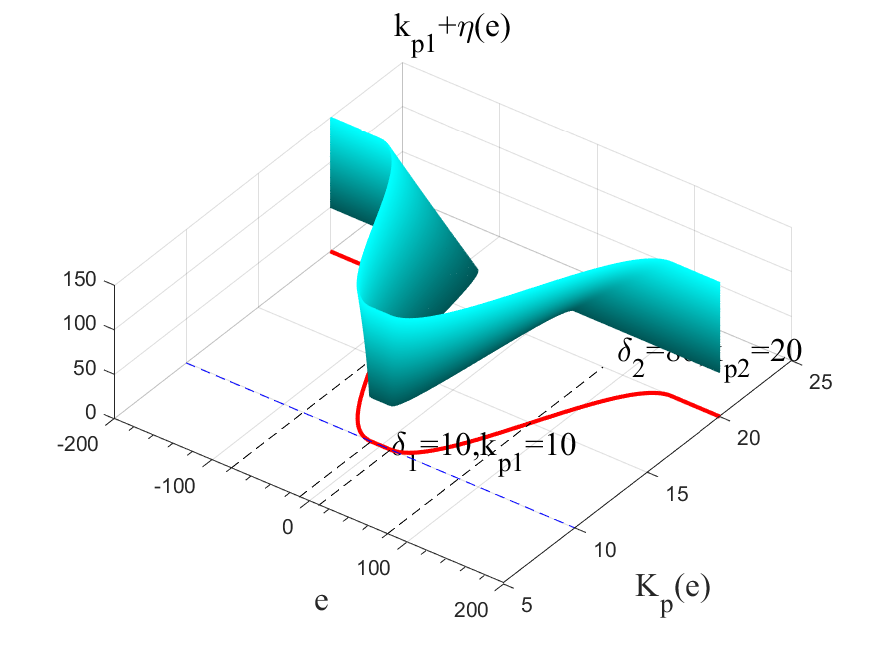}
	\caption{The change process of NLVG-PID.}
	\label{fig_0301-NLVG-PID2delta}
\end{figure}
In order to improve the PID controller's fast control capability for errors of different sizes, nonlinear gain terms are used instead of the fixed terms $K_p,K_i,K_d$ in (\ref{formula_PID_basic_law}). When the small error is close to zero, the value of the NLVG parameter reaches the lower limit $k_{\min}=[k_{p1},k_{i1},k_{d1}]$. In the case of large errors, the upper limit $k_{\max}=[k_{p2},k_{i2},k_{d2}]$ is reached, as shown in Fig. \ref{fig_0301-NLVG-PID2delta}.
Hence, we design the control law as follows:  
\begin{equation}
	U(t)_{j}^{\text{NLVG-PID}}=f_{K_p}^{j}\cdot e(t) +f_{K_i}^{j}\cdot\dot{e}(t) +f_{K_d}^{j}\cdot \int_{0}^{t } e(\tau)\,d\tau ,
	\label{formula_NLVG-PID-control-law}
\end{equation}
where
\begin{equation}
	f_{K_{p}}^{j}=\left\{\begin{array}{ll}
		k_{p 1}^{j}, & |e(t)| \in\left[0, \delta_{p1}\right), \\
		k_{p 1}^{j}+\eta_{p}^{j}, & |e(t)| \in\left[\delta_{p1}, \delta_{p2}\right], \\
		k_{p 1}^{j}+2 A_{P}^{j}, & |e(t)| \in\left(\delta_{p2},+\infty\right),
	\end{array}\right.\\
	\label{formula_3.0_NLVG-PID controller design}
\end{equation}
and
\begin{equation}
	{\eta}_{p}^{j} = -A_{P}^{j}\cdot \cos\bigg(\frac{2\pi}{\delta_{p2}-\delta_{p1}} \left|e(t)\right|- \delta_{p1}\bigg)+A_{P}^{j},
	\label{formula_np}
\end{equation}
where $\delta_{1} ,\delta_{2} \in \mathbb{R}^+$, which based on the performance metrics (\ref{formula_es_J}), as shown in Fig. \ref{fig_0301-NLVG-PID2delta}.
Likewise, the integral item is
\begin{equation}
	f_{K_{i}}^{j}=\left\{\begin{array}{ll}
		k_{i 1}^{j}, & |\dot e{(t)}| \in\left[0, \delta_{i1}\right), \\
		k_{i 1}^{j}+\eta_i^{j}, & |\dot e{(t)}| \in\left[\delta_{i1}, \delta_{i2}\right], \\
		k_{i 1}^{j}+2 A_{I}^{j}, & |\dot e{(t)}| \in\left(\delta_{i2},+\infty\right),
	\end{array}\right.
	\label{formula_3.1_NLVG-pid-ki}
\end{equation}
where
\begin{equation}
	{\eta}_i^{j} = -A_{I}^{j}\cdot \cos\bigg(\frac{2\pi}{\delta_{i2}-\delta_{i1}}\left|\dot e{(t)}\right|- \delta_{i1}\bigg)+A_{I}^{j},
	\label{formula_ni}
\end{equation}
and the differential item is
\begin{equation}
	f_{K_{d}}^{j}=\left\{\begin{array}{ll}
		k_{d 1}^{j}, & | \int_{0}^{t} e(\tau ) d \tau  | \in\left[0, \delta_{d1}\right), \\
		k_{d 1}^{j}+\eta_{d}^{j}, & | \int_{0}^{t} e(\tau ) d \tau  | \in\left[\delta_{d1}, \delta_{d2}\right], \\
		k_{d 1}^{j}+2 A_{D}^{j}, & |\int_{0}^{t} e(\tau ) d \tau  | \in \left(\delta_{d2},+\infty\right),\end{array}\right. 
	\label{formula_3.1_NLVG-pid-kd}
\end{equation}
where
\begin{equation}
	{\eta}_{d}^{j} = -A_{D}^{j}\cdot \cos \bigg(\frac{2\pi}{\delta_{d2}-\delta_{d1}} \Big\vert \int_{0}^{t} e(\tau ) d \tau\Big\vert- \delta_{d1}\bigg)+A_{D}^{j}.
	\label{formula_nd}
\end{equation}

\begin{remark}
The variables $k_{1}$ and $k_{2} = k_1+2A^j$ in (\ref{formula_3.1_NLVG-pid-kd}) are learning through the ES method {\citep{34krstic2006PID}}.
\end{remark}
Defining the state vector of the quadcopter as:  
\begin{equation}
	\begin{aligned}
		X&=\left[\begin{array}{llllllllllll}
			\phi & \dot{\phi} & \theta & \dot{\theta} & \psi & \dot{\psi}  & x & \dot{x} & y & \dot{y} & z & \dot{z}\end{array}\right]^{\top},
	\end{aligned}
	\label{formula_3.1_x}
\end{equation}
and the input vector of controller can be formulated as \begin{equation}
	\begin{aligned}
	U=&\left[U_{x}, U_{y}, U_{z}, U_{\phi}, U_{\theta}, U_{\psi}\right] \\
	=&\left[\ddot{x}_{real},\ddot{y}_{real}, F_{t}, \Gamma_{x}, \Gamma_{y}, \Gamma_{z}\right],
	\end{aligned}
\end{equation}
where $U \in \mathbb{R}^{6}$, and
\begin{equation}
	\left\{\begin{array}{l}
		U_{x}=\ddot{x}_{desire}+f_{K_p}^x\left(e_{x}\right)+f_{K_i}^x\left(\int e_{x} d t\right) +f_{K_d}^x\left(\dot{e}_{x}\right)\\
		U_{y}=\ddot{y}_{desire}+f_{K_p}^y\left(e_{y}\right)+f_{K_i}^y\left(\int e_{y} d t\right)+f_{K_d}^y\left(\dot{e}_{y}\right) \\
		U_{z}= f_{K_p}^{z} \left(e_{z}\right)+f_{K_i}^z\left(\int e_{z} d t\right)+f_{K_d}^z\left(\dot{e}_{z}\right)\\
		U_{\phi}=f_{K_p}^{\phi}\left(e_{\phi}\right)+f_{K_i}^{\phi}\left(\dot{e}_{\phi}\right)+f_{K_d}^{\phi}\left(\int e_{\phi} d t\right)\\
		U_{\theta}=f_{\theta 1}\left(e_{\theta}\right)+f_{\theta 2}\left(\dot{e}_{\theta}\right)+f_{\theta 3}\left(\int e_{\theta} d t\right)\\
		U_{\psi}=f_{\psi 1}\left(e_{\psi}\right)+f_{\psi 2}\left(\dot{e}_{\psi}\right)+f_{\psi 3}\left(\int e_{\psi} d t\right).
	\end{array}\right.
	\label{formular_control_system}
\end{equation}
where $e_i= e_{i,desire} - e_{i,real}$, and$ i \in \{x,y,z,\phi,\theta,\psi\}$. 
In (\ref{formular_control_system}), $e_{i,real}$ is the real value of $e_i$, and $e_{i,desire}$ is the desire value of $e_i$. 

\subsection{ES in NLVG-PID}
\label{subsection_ES}
In ES, disturbance driving is used to find and maintain an extremum value for dynamic systems. This approach was first proposed by Leblanc, and the stability analysis of ES was first generally implemented by Krstic in \citep{34krstic2000stability}. We used the ES method to calculate the local optimal PID gain for the controller. The input is the step signal (e.g., maximum desired angle or position), while the output is the boundary of variable gains, where $\vec {K}{_1} =(k_{p1}, k_{i1}, k_{d1})$ and $\vec {K}{_2} =(k_{p2}, k_{i2}, k_{d2})$. $\vec{K}=(k_{p}, k_{i}, k_{d})$. The cost function of drone performance during the time $t \in [t_0,t_f]$ can be defined as
\begin{equation}
	{\mathcal{J} }(\vec{K})=\frac{1}{t_{f}-t_{o}} \int_{t_{0}}^{t_{f}} e^{2}(t, \vec{K}) {d} t,
	\label{formula_es_J}
\end{equation}
where $e(t, \vec{K})$ denotes the error in following the desired path with initial disturbance using parameters $\vec{K}$. Suppose the cost function gradient, $\nabla \mathcal{J} (\vec{K})$, is available and could iteratively improve the parameter of the NLVG-PID controller. The parameter iterate rule can be defined as
\begin{equation}
	\vec{K}(t)=\vec{K}(t-1)-\alpha \nabla \mathcal{J}(\vec{K(t-1)}),
\end{equation}
where $\alpha$ denotes the step size of the parameter of the quadcopter at each step of the iteration, and $\vec{K}(t-1) $ denotes the parameter vector after $t-1$ times iterations. Then the key point is to estimate $\nabla {\mathcal{J} (\vec{K}(t-1))}$. That is
\begin{equation}
	\nabla \mathcal{J} (\vec{K})=
 \left(\frac{\partial}{\partial K_{p}} \mathcal{J} (\vec{K}), \frac{\partial}{\partial K_{i}} \mathcal{J} (\vec{K}), \frac{\partial}{\partial K_{d}} \mathcal{J} (\vec{K})\right).
\end{equation}
From (\ref{formula_es_J}), the derivative to the gains numerically can be computed as
\begin{equation}
	2\frac{\partial}{\partial K} J(\vec{K}) \approx \frac{\mathcal{J} \left(\vec{K}+\delta \cdot \hat{U}_{K}\right)-\mathcal{J} \left(\vec{K}-\delta \cdot \hat{U}_{K}\right)}{\delta},
	\label{formula_es_JK}
\end{equation}
where $U_K$ is the unit vector in the $K$ direction. When $\delta \rightarrow 0$, this approximation can minimize the cost function and optimize the local boundary PID parameters. The parameters are randomly initialized to positive numbers, and the periodic perturbation function is evaluated to evaluate the varying PID parameters and gradients of $\mathcal{\mathcal{J} } (\vec{K})$. To avoid the cost function falling into a local optimum, the initial value of the optimization value needs to be randomly reset multiple times and compared to find the best result. By plotting the functional relationship between the cost function and the number of iterations, we can observe the convergence of the system and verify the working principle of this learning method.

\subsection{Strategy planning of obstacle avoidance }
The stereo cameras are used to detect the largest threat in the visual range of the cameras, which fuse with distance from the ultrasonic sensor to the obstacles $Obj_{risky}^j$ front at the visual risk assessment models. The relative position of the quadcopter is $p = (x, y, z)$ and the direction is $v = (v_x, v_y, v_z)$. The $l_s$ denotes the distance of the current position and the boundary safety position of quadcopter $P_{safe}^i $, and the safety radius $R_{safe}$ of the generated 3D sphere $\varOmega $ in intersecting line on the direction of the velocity vector, Bessel curve fitting is carried out by setting the trajectory of UAV that meets the dynamic constraints. 
\begin{figure}[htbp]
	\centerline
	{\includegraphics[width=0.47\textwidth]{ 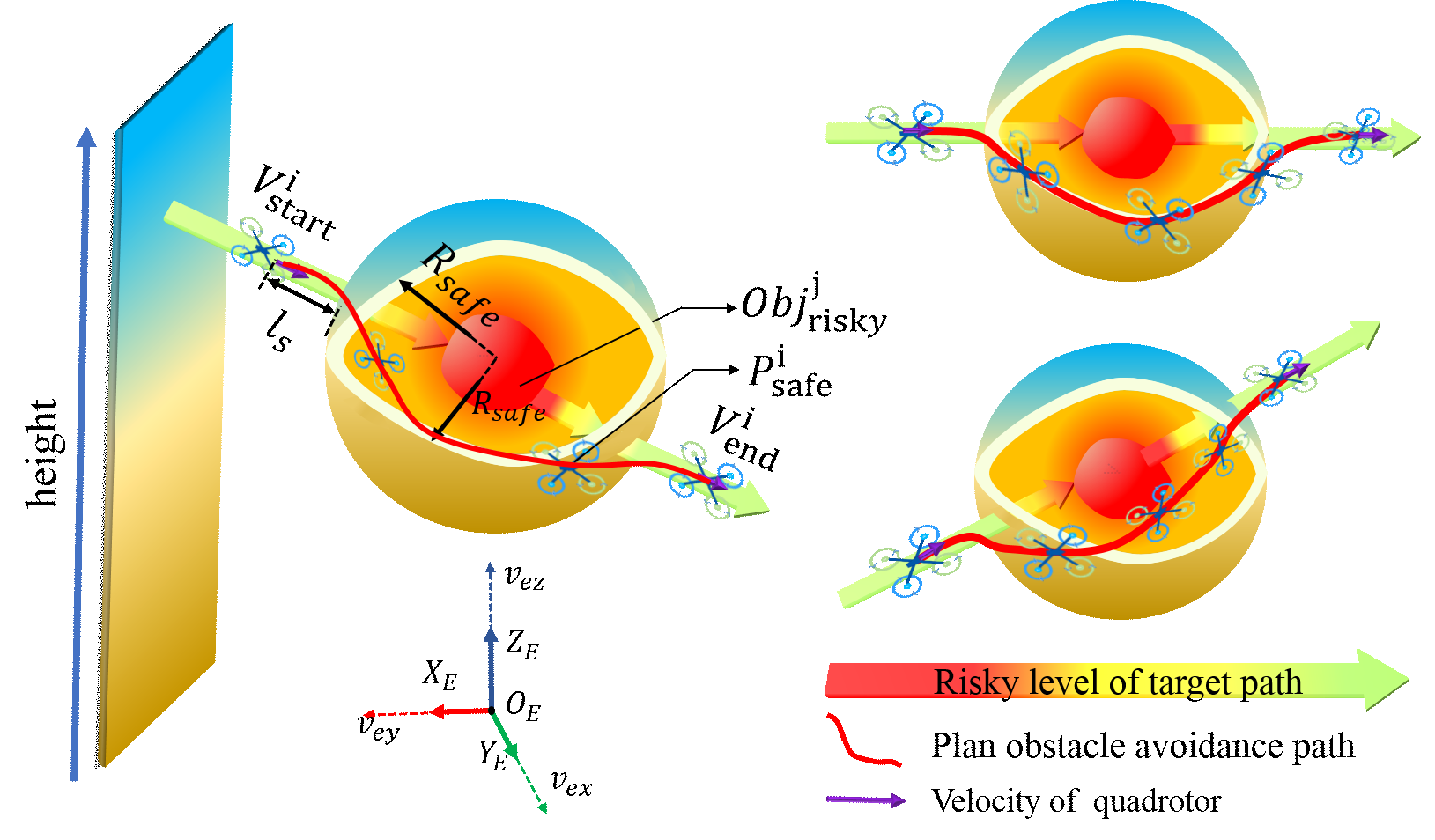}}
	\caption{Quadcopter path planning of free-collision in 3D space.}
	\label{fig_0304_pathplaning01}
\end{figure}
\section{Simulation and results}
\label{section_simulation}
\begin{figure}[htbp]
	\centerline
	{\includegraphics[width=0.45\textwidth]{ 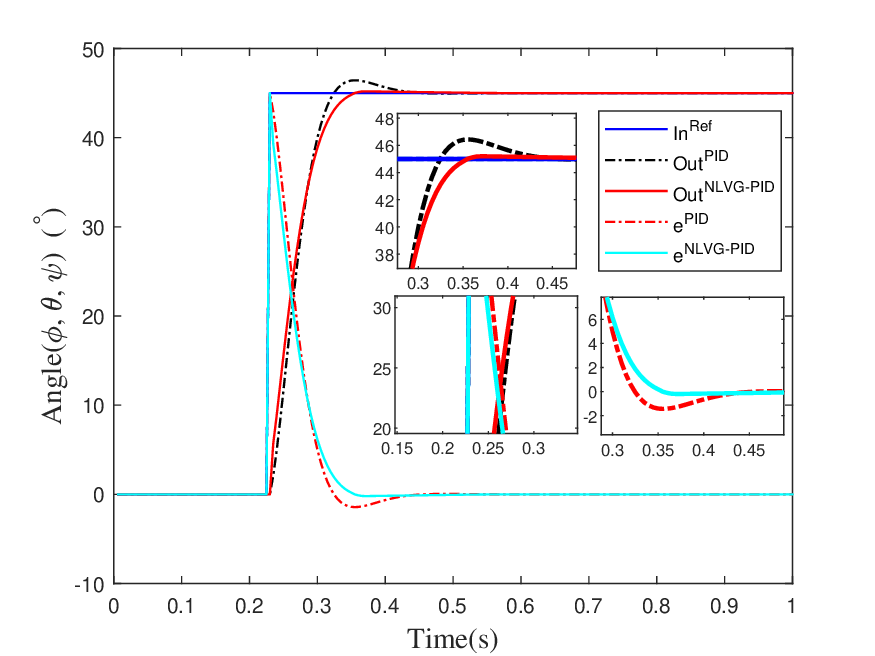}}
	\caption{Attitude controller step-response of PID and NLVG-PID}
	\label{fig_0304_pid-vs-nlvgpid-v02.eps}
\end{figure}

A simulation experiment was established to verify the efficiency of the NLVG-PID controller. The specific test parameters are set as \citep{yanhui2022CCC}.
The rate inner loop controller is the lowest-level controller of the quadcopter. Aiming at the internal loop angle control's step response, PID control and NLVG-PID plus ES control simulation experiments were conducted. The results and analysis are shown in Fig. \ref{fig_0304_pid-vs-nlvgpid-v02.eps} and Table \ref{Table_Performance_PID_NLVG}. It should be noted that the performance of the NLVG-PID controller uses three cost functions (IAE: integrated absolute error, ITAE: integration time and absolute error, ITSE: integrated time-weighted squared error), as set $t^{angle}_{peak1}=0.15s$, $t^{position}_{peak2}=0.37s$.

\begin{table*}[thp]\footnotesize
	\centering
        \label{Table_Performance_PID_NLVG}
	\caption{Results for fixed-PID and NLVG controller in step singal of quadcopter  } 

	\addtolength{\tabcolsep}{4.8pt}
	\begin{tabular*}{16.4cm}{ccccccccc}
		\toprule[0.75pt]
		\multirow{2}{*}{Channel}   & \multicolumn{3}{c}{PID}  &    & \multicolumn{3}{c}{NLVG-PID (Ours)}& \multicolumn{1}{c}{Function} \\
		\cmidrule[0.5pt]{2-4} 
		\cmidrule[0.5pt]{6-8} & IAE & ITAE & ITSE &    & IAE & ITAE & ITSE  &-\\
		\midrule[0.5pt]
		$\phi$    &1025.01  &5.10  & 185.75&    &933.8 &4.71  &156.42 & attitude in x axis ($^o$)\\
		$\theta$  &1015.29  &5.03  & 172.42&    &910.0 &4.33  &140.90 & attitude in y axis ($^o$)\\
		x       &94.82    &0.47    &0.92  &    &86.32    &0.43    &0.78 & position x\\
		y       &91.82    &0.41    &0.89  &    &82.66    &0.36    &0.70 & position y\\
		\bottomrule[0.75pt]
		\multicolumn{9}{p{16.6cm}}{$^*$
			$\mathrm{IAE}=\frac{1}{t_{peak}} \int_{0}^{t_{peak}} |e| d t$;
			$\mathrm{ITAE}=\frac{1}{t_{peak}} \int_{0}^{t_{peak}} t|e| d t$; $\mathrm{ITSE}=\frac{1}{t_{peak}} \int_{0}^{t_{peak}} t e^{2} d t$.}
	\end{tabular*}
\end{table*}
The initial parameters of the quadcopter can be determined as \citep{52yanhui2020CAC}. First, the inner loop controller parameters $K_{PID}^{\phi}={(8, 0.1, 5)}$, $K_{PID}^{\theta}={(8, 0.1, 5)}$, $K_{PID}^{\psi}={(8, 0.1, 5)}$, and the outter loop position controller initial parameters are $K_{PID}^{\phi}={(5, 0.2, 5)}$, $K_{PID}^{\theta}={(5, 0.2, 5)}$, $K_{PID}^{\psi}={(5, 0.2, 5)}$. Second, the scaler gains $\delta_1=0.01 $, $\delta_2 =0.838$, then the controller parameters are learned by ES in subsection \ref{subsection_ES}. Moreover, the step size of this simulation is $\Delta t=0.01s$; the total time is $T_{total} =140s$. 

\subsection{Example 1: Storm path following}
\label{Example 1}
\begin{figure}[htbp]
	\centering
	\subfigure{a) Storm-type obstacle avoidance routes in 3D view.}{
		{\includegraphics[width=0.50\textwidth]{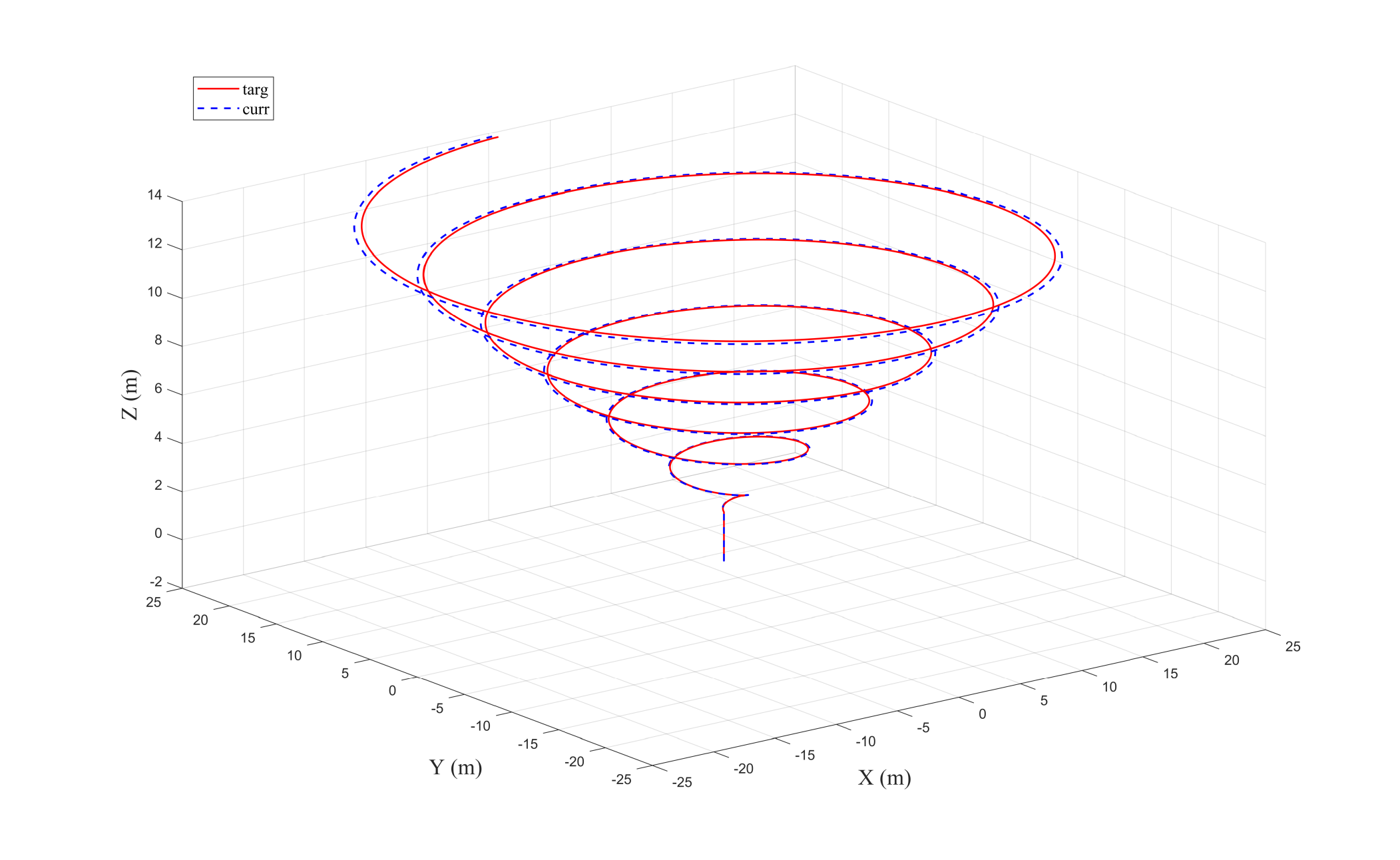}}
	}
	\subfigure{b) Storm-type obstacle avoidance routes in X-Y view.}{
		{\includegraphics[width=0.44\textwidth]{ 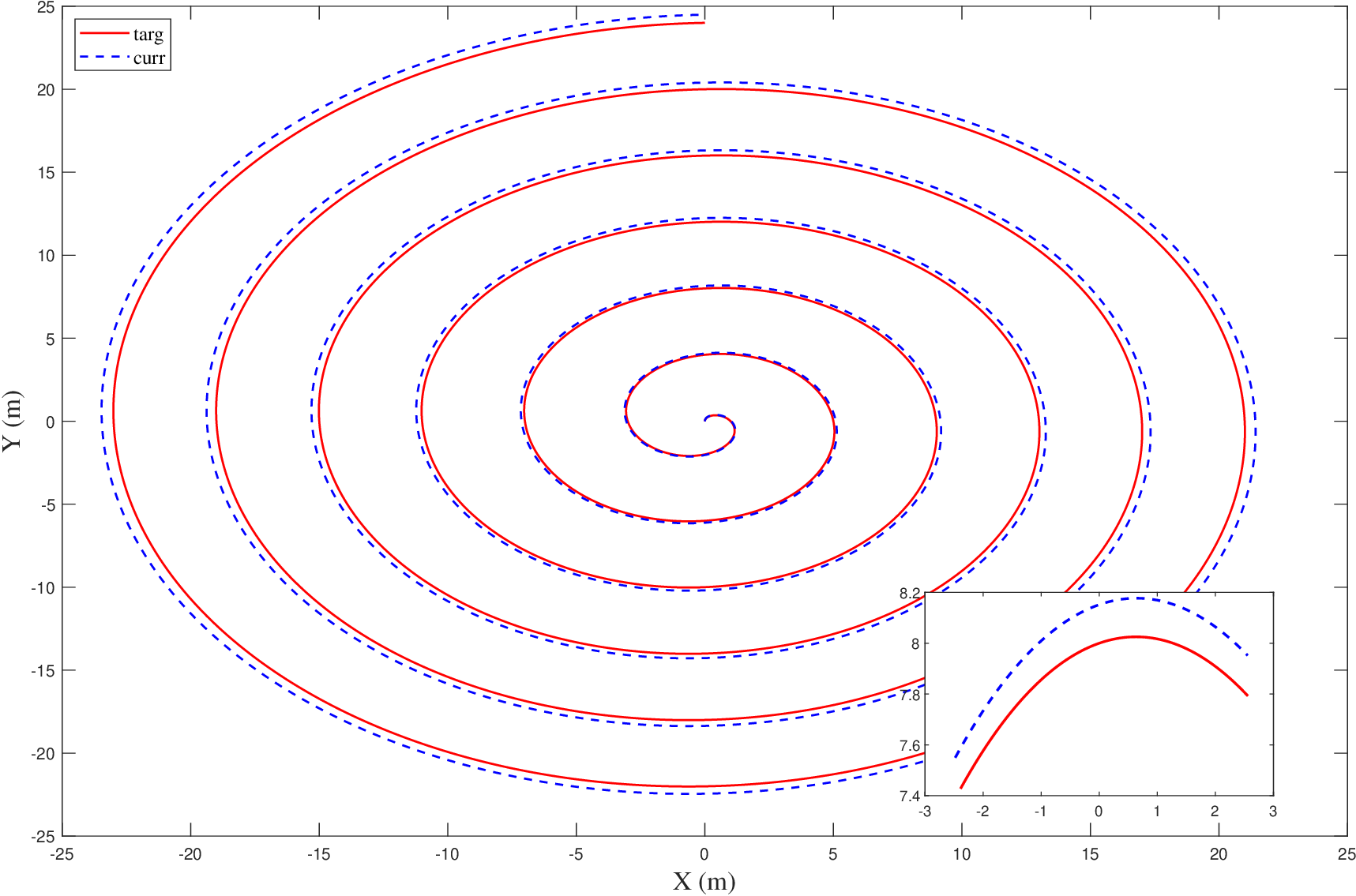}}
	}
	\caption{Storm-type obstacle avoidance routes}
	\label{fig_Storm_type_obstacle_avoidance_route}
\end{figure}

To verify the effect of large-angle turning and small-angle adjustment of aircraft under the disturbance of obstacles, a storm airline is a continuous arc of increasing radius and height, designed as Fig. \ref{fig_Storm_type_obstacle_avoidance_route}.

\begin{figure}[htbp]
	\centerline
	{\includegraphics[width=0.47\textwidth]{ 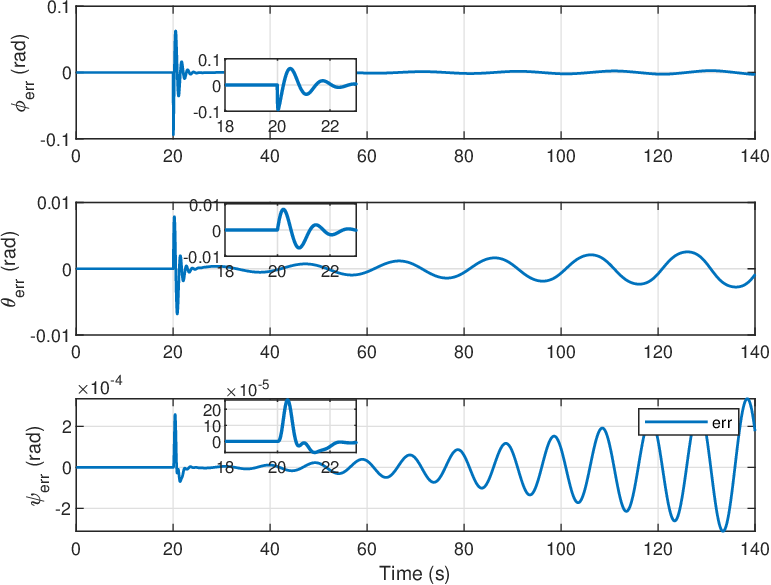}}
	\caption{PID: angle error of Storm-type path.}
	\label{fig_0304_20201219-PID-storm-err-angle}
\end{figure}

\begin{figure}[htbp]
	\centerline
	{\includegraphics[width=0.47\textwidth]{ 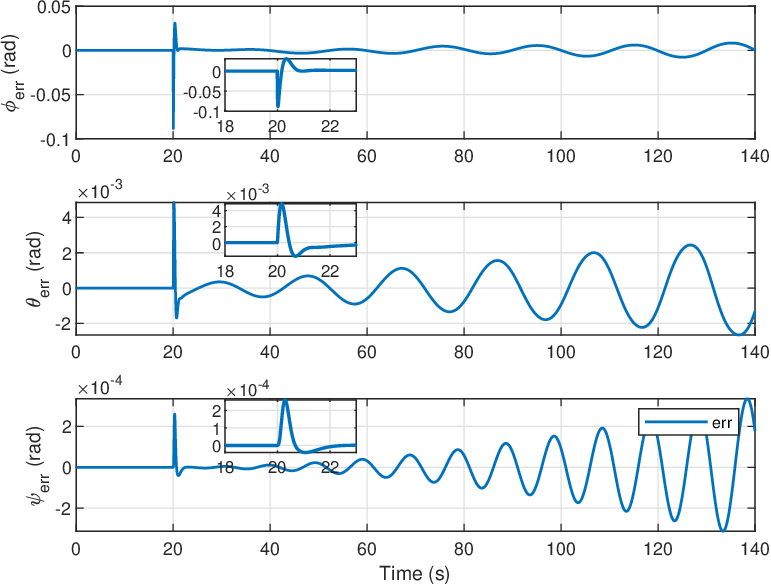}}
	\caption{NLVG-PID: angle error of Storm-type path.}
	\label{fig_0304_NLVG-PID-storm-err-angle}
\end{figure}

\begin{figure}[htbp]
	\centerline
	{\includegraphics[width=0.47\textwidth]{ 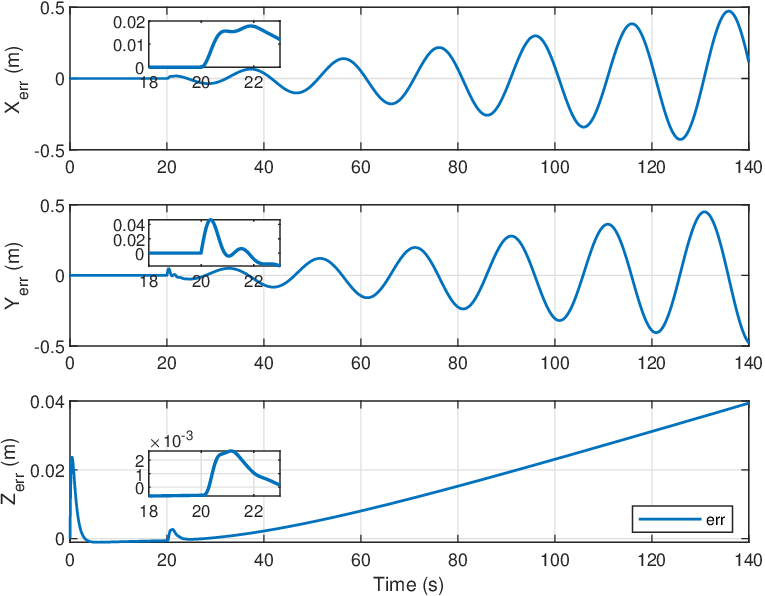}}
	\caption{PID: position error of the Storm-type path}
	\label{fig_0304_PID-storm-err-xyz}
\end{figure}

As shown in the Fig. \ref{fig_Storm_type_obstacle_avoidance_route}-\ref{fig_0304_PID-storm-err-xyz}, when the aircraft executes the takeoff command at t=20s, the attitude angle error is significant at this time. Then, PID and dynamic adaptive control algorithms are used for individual control. During the execution of the storm path, the attitude angle error decreases, and the control optimization effect is significantly improved.
The Euler angle $(\phi, \theta, \psi)$, the angle error $(\phi_{err}, \theta_{err}, \psi_{err})$, the position $(X, Y, Z)$, and the position error $(X_{err}, Y_{err}, Z_{err})$ in OATC are described in Fig. \ref{fig_Storm_type_obstacle_avoidance_route}-\ref{fig_0304_PID-storm-err-xyz}, respectively. From these figures, it can be observed that the angle and position errors can lead to a fast convergence into the local neighborhood areas in both the fixed-gain PID controller and NLVG-PID. However, the proposed NLVG-PID with ES control systems can perform well in OATC missions. It should be noted that the altitude of these missions is to keep the same value as $10m$.

\begin{figure}[htbp]
	\centering
	\subfigure{a) Lissajous-type path in 3D view.}{
		{\includegraphics[width=0.49\textwidth]{ 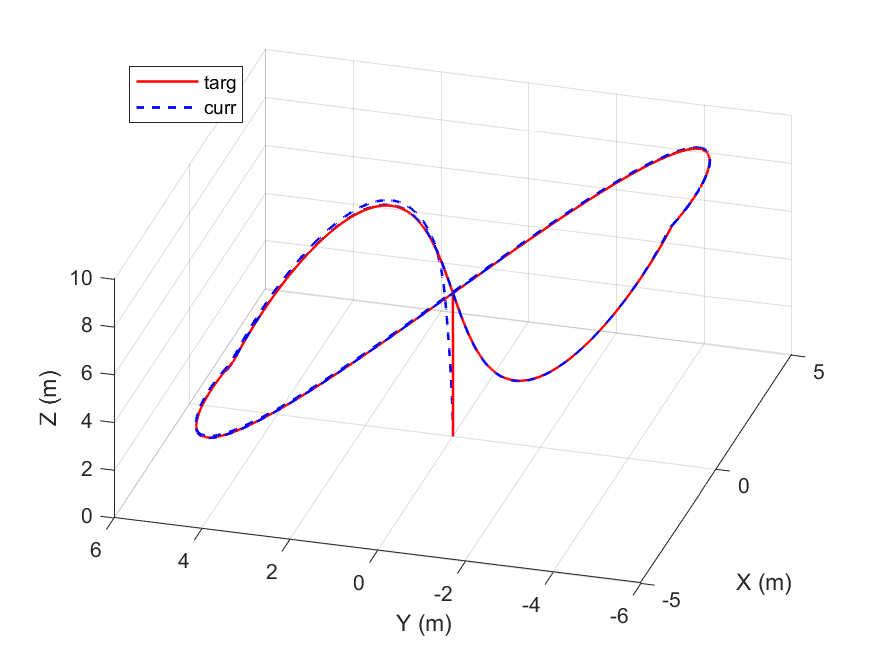}}
	}
	\subfigure{b) Lissajous-type path in X-Y view.}{
		{\includegraphics[width=0.47\textwidth]{ 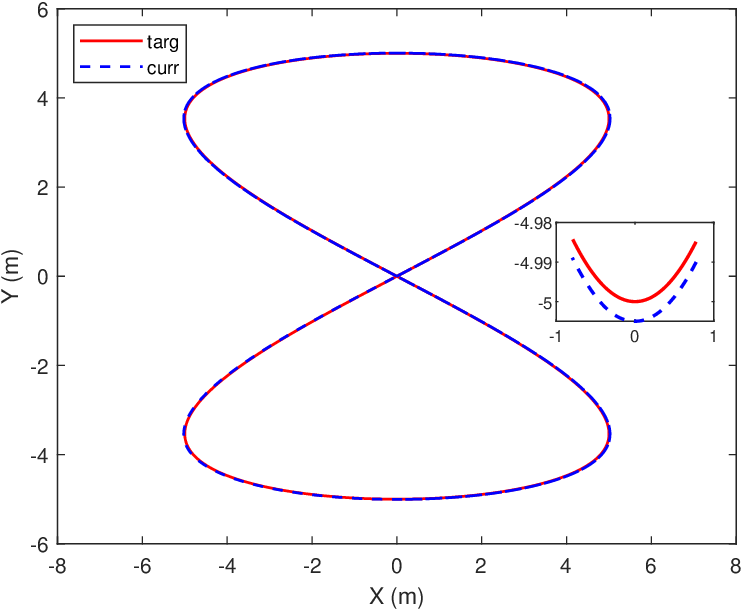}}
	}
	\caption{Lissajous-type path following.}
	\label{Lissajous-type path following}
\end{figure}

\begin{figure}[htbp]
	\centerline
	{\includegraphics[width=0.47\textwidth]{ 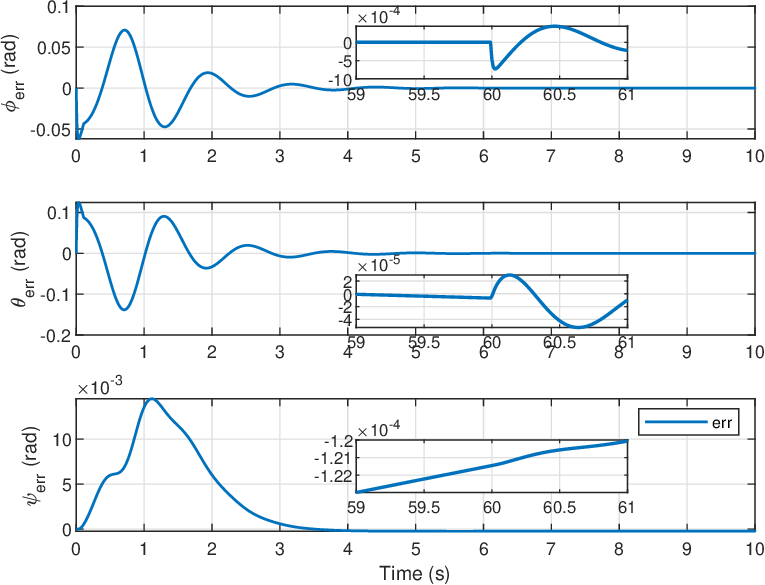}}
	\caption{PID: attitude angle error in  Lissajous-type path.}
	\label{fig_0304_20201219-PID-eight-err-angle}
\end{figure}

\begin{figure}[!thp]
	\centerline
	{\includegraphics[width=0.47\textwidth]{ 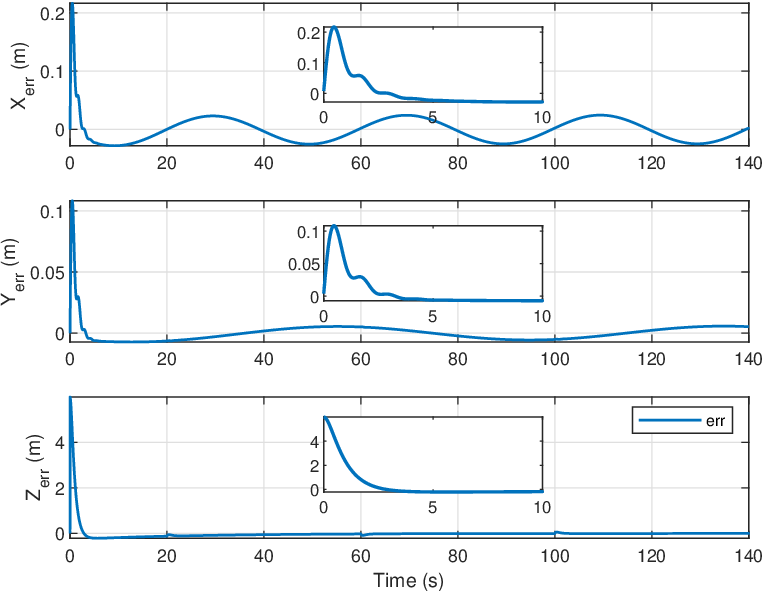}}
	\caption{PID: position errors in Lissajous-type path.}
	\label{fig_0304_PID-Eight-err-xyz}
\end{figure}

\begin{figure}[!thp]
	\centerline
	{\includegraphics[width=0.47\textwidth]{ 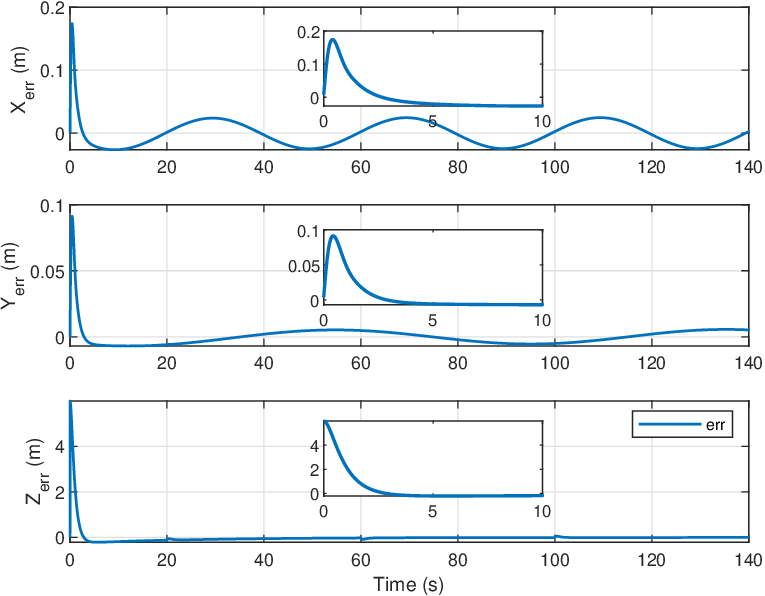}}
	\caption{NLVGPID: position errors in Lissajous-type path.}
	\label{fig_0304_NLVG-PID3-Eight-err-xyz}
\end{figure}

The desired trajectory path (Lissajous-type) of the quadcopter is depicted in Fig. \ref{Lissajous-type path following}, and the Euler angle and the position error in OATC are described in Fig. \ref{Lissajous-type path following}-\ref{fig_0304_NLVG-PID3-Eight-err-xyz}, respectively. As can be seen from these figures, using NLVG is better than PID in calculating a fast and smooth control output, thereby quickly tracking the desired command. It is worth noting that the height of this mission dynamically changes with the Lissajous-type path, which can verify that the proposed NLVG-PID performs well in periodic 3D path tracking.

\subsection{Example 2: Lissajous curve following}
\label{Example 2}
This work designed a 3D-Lissajous curve to conduct flight simulation verification of the quadcopter's control performance under periodic disturbances.
\begin{figure}[htbp]
	\centerline
	{\includegraphics[width=0.47\textwidth]{ 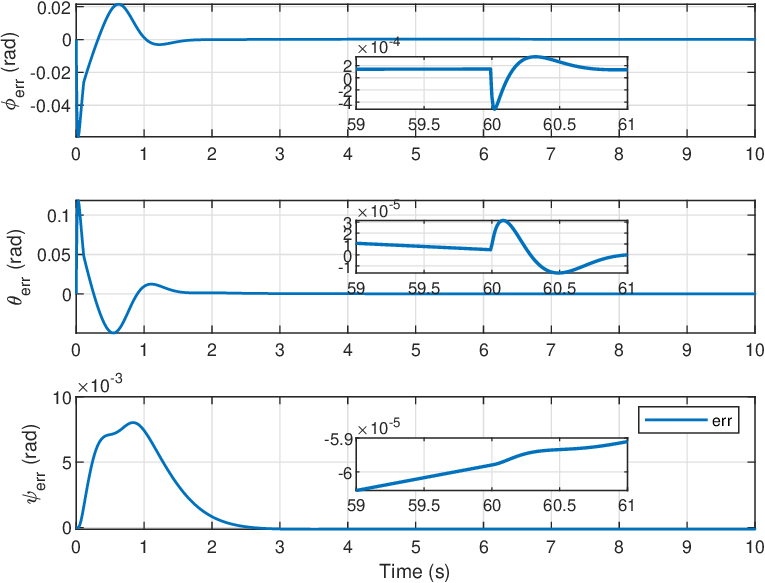}}
	\caption{NLVG-PID: attitude angle error in  Lissajous-type path.}
	\label{fig_0304_NLVG-PID-Eight-err-angle}
\end{figure}


As can be seen in Fig. \ref{fig_0304_NLVG-PID-Eight-err-angle}, when the aircraft executes the takeoff command at t=0s, the position and attitude angle errors at the moment of takeoff are significant. Then, PID and dynamic adaptive control algorithms are used for control, respectively. During the execution of the Lissajous path, the attitude angle error is gradually reduced by about 50\%, the position error is reduced by about 15\%, and the control optimization effect is partially improved. Compared with Example \ref{Example 1}, the path of Example \ref{Example 2} is more complex, and there are significant changes in height and position, so the effect improvement is slightly reduced.
\section{Conclusion}
\label{section_conclusion}
This paper studies the active motion control problem of quadcopter obstacle avoidance. A new design scheme for adaptive learning control of flight controllers based on low-cost dynamic linear optimization under uncertain conditions with obstacles is proposed. First, an NLVG-PID controller is presented for the formulated UAV model. Furthermore, ES is used to learn optimal NLVG-PID parameters in offline maximum or minimum error step signals. Numerical simulations were performed based on this design, and the results show that the proposed adaptive learning controller can reduce response overshoot and settling time in typical 3D path curves (such as storm paths).


\end{document}